\documentclass[useAMS,usenatbib,usegraphicx]{mn2e}
\usepackage{amsmath,euscript,amsfonts,amssymb}
\usepackage{times}
\usepackage[colorlinks=True, citecolor=blue, menucolor=green, linkcolor=red]{hyperref}
\usepackage[all]{hypcap}
\usepackage[dvipsnames]{xcolor}
\usepackage{multirow}

\def\ergps{erg s$^{-1}$}
\def\ergphzps{erg Hz$^{-1}$ s$^{-1}$}
\def\Msunpyr{M$_{\odot}\,$yr$^{-1}$}

\def\mJy{$\mu$Jy}
\def\kmps{km s$^{-1}$}
\def\araa{ARA\&A}
\def\aap{A\&A}
\def\nat{Nature}

\def\mnras{MNRAS}
\def\apj{ApJ}

\def\apjl{ApJL}
\def\pasp{PASP}

\newenvironment{figurehere*}
  {\def\@captype{figure*}}
  {}
\makeatother

\title[SN 2007bg: The Complex Circumstellar Environment]{SN 2007bg: The Complex
Circumstellar Environment Around One of the Most Radio-Luminous Broad-Lined Type Ic Supernovae}

\author[Salas et al.]{P. Salas,$^{1}$ F. E. Bauer,$^{1,2}$ C.
Stockdale$^{3,4}$ J. L. Prieto,$^{5}$\thanks{Hubble, Carnegie-Princeton Fellow.}\\
$^{1}$Pontificia Universidad Cat\'olica de Chile, Departamento de Astronom\'ia y
Astrof\'isica, Casilla 306, Santiago 22, Chile\\
$^{2}$Space Science Institute, 4750 Walnut Street, Suite 205, Boulder, CO 80301\\
$^{3}$Marquette University, Physics Department, PO Box 1881,
Milwaukee, WI 53201, USA\\
$^{4}$The Homer L. Dodge Department of Physics \& Astronomy, 440 W. Brooks ST,
The University of Oklahoma, Norman, OK 73019, USA\\
$^{5}$Department of Astrophysical Sciences, Princeton University, Peyton Hall,
Princeton, NJ 08544, USA\\
}

\begin{document}

\date{\today}

\pagerange{\pageref{firstpage}--\pageref{lastpage}} 
\pubyear{2012}

\maketitle
\label{firstpage}

\begin{abstract}

In this paper we present the results of the radio light curve and X-ray
observations of broad-lined Type Ic SN 2007bg. The light curve shows three distinct phases of
spectral and temporal evolution,
implying that the SNe shock likely encountered at least 3 different circumstellar medium regimes. We
interpret this as the progenitor of SN 2007bg having at least two distinct mass-loss episodes (i.e.,
phases 1 and 3) during its final stages of evolution, yielding a highly-stratified circumstellar
medium.
Modelling the phase 1 light curve as a freely-expanding, synchrotron-emitting shell, self-absorbed
by its own radiating electrons, requires a progenitor mass-loss rate of
$\dot{M}\approx1.9\times10^{-6}(v_{w}/1000\;\rm{km\,s}^{-1})$ \Msunpyr\ for the last
$t\sim20(v_{w}/1000\;\rm{km\,s}^{-1})$ yr
before explosion, and a total energy of the radio emitting ejecta of
$E\approx1\times10^{48}$ erg after $10$ days from explosion. This places SN 2007bg among
the most energetic Type Ib/c events. We interpret the second phase as a sparser "gap" region between
the two winds stages. Phase 3 shows a second absorption turn-on before rising to a peak luminosity
$2.6$ times higher than in phase 1. Assuming this luminosity jump is due to a circumstellar medium
density enhancement from a faster previous mass-loss episode, we estimate that the phase 3 mass-loss
rate could be as high as $\dot{M}\la4.3\times10^{-4}(v_{w}/1000\;\rm{km\,s}^{-1})$ \Msunpyr. The
phase 3 wind would have transitioned directly into the phase 1 wind for a wind speed difference of
$\approx2$. In summary, the radio light curve provides robust evidence for dramatic global changes
in at least some Ic-BL progenitors just prior ($\sim 10-1000$ yr) to explosion.
The observed luminosity of this SN is the highest observed for a non-gamma-ray-burst 
broad-lined Type Ic SN, reaching $L_{8.46\,\rm{GHz}}\approx1\times10^{29}$ \ergphzps,
$\sim567$ days after explosion.

\end{abstract}


\section{Introduction}

Core collapse supernovae (SNe) mark the spectacular death of massive
stars \citep*{1986ARA&A..24..205W}. Among these SNe Types Ib and Ic (collectively referred to as
Ib/c due to their perceived similarities) are associated with the explosion of massive stars which
were stripped of their H envelopes by strong stellar winds. Types Ib show He features in their
spectra, whereas Types Ic do not. Our current understanding of Type Ib/c SNe suggests that
Wolf-Rayet (WR) stars are a primary progenitor path \citep*{2003PASP..115....1V}, although the
relative numbers of WR stars fall well short of the observed Type Ib/c rates and thus a lower-mass
binary progenitor path is most likely required to account for a significant fraction of
these events \citep[e.g.,][]{1992ApJ...391..246P,2004MNRAS.349.1093R,2011ApJ...740...79W,Smith2011}.
Archival imaging of pre-SNe locations, however, have failed to produce direct
associations between Type Ib/c SNe and their possible progenitors
\citep[e.g.,][]{2009ARA&A..47...63S}. Thus we must rely both
on theory and indirect information obtained after the explosion. Fortunately,
X-ray and radio emission are detectable by-products of the shock interaction
between the high-velocity ejecta and the low-velocity progenitor wind in SNe
(whereas the optical emission is generated by lower-velocity ejecta and
radioactive decay). These observables crudely scale with the density of the
circumstellar medium (CSM), and hence provide constraints on the wind properties of the SN
progenitor \citep[e.g.,][]{1982ApJ...258..790C,1982ApJ...259..302C}. Because massive stars
are expected to have strong episodic mass-loss \citep[e.g.,][]{1994PASP..106.1025H} and line driven
winds \citep*[e.g.,][]{Vink2000}, radio monitoring of Type Ib/c SNe provide interesting constraints
on possible progenitor types.
The past decade has witnessed increased attention on Type Ib/c SNe due to the
association of Type Ic SN 1998bw with gamma-ray burst (GRB) 980425
\citep{1998Natur.395..670G}, as these objects may help elucidate how the central
engines of both GRBs and their associated SNe work. Following this historic connection, several
other Type Ic SNe have been linked to GRBs
\citep[e.g.,][]{Hjorth2003,Kawabata2003,Matheson2003,2003ApJ...591L..17S,2005ApJ...619..994F,
2006ApJ...645L..21M,2006ARA&A..44..507W,Bufano2012};  all of these objects show broad
emission lines in their optical spectra, with velocities above $15000$ \kmps. To date so far 5 GRBs
have been spectroscopically confirmed to have a SNe counterpart, while 14 show photometric
evidence of a SNe counterpart \citep*{Hjorth2011}, and hence a larger sample is needed in order to
shed some light on the explosion mechanisms involved. Given the large number of SNe Type
Ic compared to GRBs, one key question is whether these SNe could simply be off-axis GRBs.
Initial efforts concluded that the vast majority were not \citep[e.g.,][]{2006ApJ...638..930S}.
Recently new evidence was presented for the detection of at least one possible off-axis
GRB with no direct observation of the gamma-ray emission: SN 2009bb \citep{2010Natur.463..513S};
as well as the more controversial SN 2007gr
\citep{2010Natur.463..516P,2010ApJ...725..922S,2011ApJ...735....3X}.

Here we report on a particularly interesting broad-lined Type Ic, SN 2007bg, which was optically
discovered on 2007 April 16.15 UT at $\alpha=11^{\rm h} 49^{\rm m} 26.18^{\rm s}$,
$\delta=+51^{\circ} 49^{\prime} 21.8^{\prime\prime}$ (J2000) at redshift 
$z=0.0346$ \citep*{2007CBET..927....1Q} ($d=152$ Mpc, $H_{0}=70$ km s$^{-1}$
Mpc$^{-1}$, $\Omega_{\rm{M}}=0.3$, and $\Omega_{\rm{\Lambda}}=0.7$ in a $\Lambda$CDM cosmology).
A spectrum taken on 2007 April 18.3 with the 9.2 m Hobby-Everly Telescope let
\citet{2007CBET..927....1Q} classify SN 2007bg as a Type Ic broad-lined (Ic-BL) supernova.
The classification of this SN as a Type Ic and the fact that it
resides in a faint host encouraged \citet*{2008ApJ...673..999P} to think about
this object as a likely off-axis GRB. \citet*{2009ATel.2065....1P} also noted
that SN 2007bg was one of the brightest radio SNe one year after explosion,
making it an even better candidate for an off-axis GRB. However, based on the
radio light curve of SN 2007bg known at the time, \citet{2009ATel.2066....1S} contended that the
observed radio emission could be explained by the presence of a density enhancement
in the CSM, rather than a GRB like central engine driving the explosion. Using
the archival radio data for SN 2007bg, we further address this issue below.

The article outline is the following.
In \S\ref{s:obs} we introduce the observations taken with the Very Large Array (VLA). Our results
and modelling of the radio light curves are presented in \S\ref{s:results}. Finally in
\S\ref{s:discussion} we summarise our findings and provide a brief discussion on the future of radio
observations of SNe.

\section{Observations}
\label{s:obs}

\subsection{Radio}

The VLA\footnote{The Very Large Array and Very Long Baseline Array are operated by the National
Radio Astronomy Observatory, a facility of the National Science Foundation operated under
cooperative agreement by Associated Universities, Inc.}
radio observations were carried out between 2007 April 19 and 2009 August 28 as part of the programs
AS887, AS929 and AS983 (PI A. M. Soderberg).

The data were taken in standard continuum observing mode, with a bandwidth of $2\times50$ MHz.
Data were collected at $1.43$, $4.86$, $8.46$, $15$, $22.5$ and $43.3$ GHz.
As primary flux density calibrators 3C286 and 3C147 were used. We used these calibrators to
scale our flux density measurements to the Perley-Buttler 2010 absolute flux scale.
For secondary calibrators three different sources were used; SDSS J114856.56+525425.2 in most
epochs, as well as SDSS J114644.20+535643.0 and SBS 1150+497. The derived flux density for the
secondary calibrators is shown in Table \ref{tab:phase}.
Traditionally one has been able to cross-check the derived secondary calibrator fluxes with archival
data, but in this case the VLA stopped regular monitoring of calibrators since 2007. This, in
combination with the weak calibrator for the $15$ and $22.5$ GHz
data\footnote{\url{http://www.vla.nrao.edu/astro/calib/manual}}, makes this data less
reliable than the rest at lower frequencies, even adopting larger error bars for it.

\begin{table*}
 \centering
 \begin{minipage}{84mm}
  \caption{\protect{VLA secondary calibrator measurements$^{\mathsf{a}}$} \label{tab:phase}}
  \begin{tabular}{lccccc}
  \hline
  \hline
  Obs. & $f_{1.43\,\rm{GHz}}$ & $f_{4.86\,\rm{GHz}}$ & $f_{8.46\,\rm{GHz}}$ & $f_{15\,\rm{GHz}}$ &
$f_{22.5\,\rm{GHz}}$\\
  Date & (Jy)     & (Jy)    & (Jy)      & (Jy)    & (Jy) \\
  \hline
  2007-Apr-19 &        &        & 0.4525 &        &           \\
  2007-Apr-23 &        &        & 0.4306 &        & 0.2732    \\
  2007-Apr-24 &        & 0.4571 & 0.433 &        &           \\
  2007-Apr-25 &        &        & 0.453 &  & 0.282 \\
  2007-Apr-26 &        & 0.4552 & 0.4521 & 0.3644 &           \\
  2007-Apr-30 &        & 0.467  & 0.4583 &        &           \\
  2007-May-3  &        &        &        & 0.355 & 0.2752    \\
  2007-May-5  &        & 0.458  & 0.4427 &        &           \\
  2007-May-12 &        & 0.412  & 0.396 & 0.3607 & 0.2733    \\
  2007-May-17 &        & 0.4582 & 0.4623 & 0.3590 & 0.268     \\
  2007-May-27 & 0.4533$^{\mathsf{b}}$ & 0.4507 & 0.4173 & 0.345  & 0.2784    \\
  2007-Jun-11 &        & 0.45 & 0.464  & 0.359  & 0.2781   \\
  2007-Jun-22 &        & 0.4587 & 0.4372 &        &           \\
  2007-Jul-3  &        &        &        & 0.30 & 0.29    \\
  2007-Jul-7  &        & 0.4599 & 0.4479 &        &          \\
  2007-Jul-24 & 0.465$^{\mathsf{b}}$  & 0.453 & 0.4462 & 0.349  & 0.266     \\
  2007-Aug-18 &        & 0.4545 & 0.4410 & 0.348  & 0.2720    \\
  2007-Sep-7  & 0.465$^{\mathsf{b}}$  & 0.464  & 0.4475 & 0.352  & 0.30      \\
  2007-Sep-23 & 0.476$^{\mathsf{b}}$  & 0.4668 & 0.4624 &        &           \\
  2007-Oct-23 & 0.459$^{\mathsf{b}}$  & 0.4514 & 0.4410 &        &           \\
  2007-Nov-17 & 0.4562$^{\mathsf{b}}$ & 0.4332 & 0.4449 &        &           \\
  2007-Dec-23 & 0.452$^{\mathsf{b}}$  & 0.4583 & 0.4513 &        &           \\
  2008-Jan-3  &        &        &        &        & 0.2685     \\
  2008-Jan-28 &        & 0.50 & 0.475 &  & 0.10 \\
  2008-Feb-25 &        & 0.4642 & 0.4561 &        & 0.304     \\
  2008-Apr-19 & 0.476$^{\mathsf{b}}$  & 0.6247$^{\mathsf{b}}$ & 0.587$^{\mathsf{b}}$  &        &    
       \\
  2008-May-7  &        &        & 0.449  &        &            \\
  2008-Jun-9  &        & 1.197$^{\mathsf{c}}$  & 1.05$^{\mathsf{c}}$   &        &
1.050$^{\mathsf{c}}$      \\
  2008-Nov-3  & 0.5160$^{\mathsf{b}}$ & 0.6480$^{\mathsf{b}}$ & 0.6069$^{\mathsf{b}}$ &        &
0.491$^{\mathsf{b}}$      \\
  2008-Dec-30 &        &  & 0.481 & 0.351 &            \\
  2009-Apr-6  & 0.5100$^{\mathsf{b}}$  & 0.575$^{\mathsf{b}}$  & 0.5430$^{\mathsf{b}}$ &        &
0.3491$^{\mathsf{b}}$     \\
  2009-May-31 & 0.503$^{\mathsf{b}}$  & 0.5763$^{\mathsf{b}}$ & 0.523$^{\mathsf{b}}$  &        &    
       \\
  2009-Aug-27 & 0.530$^{\mathsf{b}}$  & 0.5411$^{\mathsf{b}}$ & 0.5162$^{\mathsf{b}}$ &        &    
       \\
  \hline
  \multicolumn{6}{l}{$^{\mathsf{a}}$ Unless otherwise stated, the flux density measurements are}\\ 
  \multicolumn{6}{l}{for SDSS J114856.56+525425.2.} \\
  \multicolumn{6}{l}{$^{\mathsf{b}}$ Flux density measurements are for SDSS J114644.20+535643.0.} \\
  \multicolumn{6}{l}{$^{\mathsf{c}}$ Flux density measurements are for SBS 1150+497.} \\
  \multicolumn{6}{l}{On 2008 January 5 the source was observed at $43.3$ GHz.}\\ 
  \multicolumn{6}{l}{The secondary calibrator flux during this epoch is $0.209$ Jy.}
\end{tabular}
\end{minipage}
\end{table*}

During this period of time, the VLA incorporated the new EVLA correlator to some of the antennas, so
some of the antennas did not have all of the different wavelength receivers. This greatly reduced
the sensitivity of observations during some epochs, especially at $1.43$ and $15$ GHz.
We flagged, calibrated, and imaged the observations using the
{\sc CASA}\footnote{\url{http://casa.nrao.edu/}} (Common Astronomy Software
Applications) software following standard procedures. We combined VLA and EVLA baselines, and to
reduce the effects of the differences between VLA and EVLA antennas, we derived baseline-dependent
calibrations on the primary flux calibrator.

To measure the flux density of the source we fitted a Gaussian at the source position in our
cleaned images, as for upper limits these were computed from the dirty map {\it rms} in the central
region.
Since the {\it rms} noise in each cleaned map only provides a lower limit on the total error we
added a systematic uncertainty $\epsilon_{\nu}$ due to possible inaccuracies of the VLA flux density
calibration and deviations from an absolute flux density scale resulting in a final error
for each flux density measurement of 
$\sigma^{2}=(\epsilon_{\nu}S_{\nu})^{2}+\sigma_{\it{rms}}^{2}$, with
$\epsilon_{\nu}=0.2,0.2,0.1,0.05,0.05,0.05$ at $43.3$, $22.5$, $15$, $8.46$, $4.86$, and $1.43$
GHz. The factors were chosen based on the quality of the secondary calibrator, and on the
observed phase scatter after calibration of the data.
The resulting flux density measurements of SN 2007bg are shown in Table \ref{tab:flux} along with
the image {\it rms}.

\begin{table*}
 \centering
 \begin{minipage}{150mm}
 \caption{\protect{VLA flux density measurements for SN 2007bg} \label{tab:flux}}
 \begin{tabular}{lcccccccc}
  \hline
  \hline
  Obs. & $\Delta t$             & $f_{1.43\,\rm{GHz}}$ & $f_{4.86\,\rm{GHz}}$ & $f_{8.46\,\rm{GHz}}$
& $f_{15\,\rm{GHz}}$ & $f_{22.5\,\rm{GHz}}$ & $f_{43.3\,\rm{GHz}}$ & Array\\
  Date &  (Days)$^{\mathsf{a}}$ & ($\mu$Jy) & ($\mu$Jy) & ($\mu$Jy) & ($\mu$Jy) & ($\mu$Jy) &
($\mu$Jy) & Configuration\\
  \hline
2007-Apr-19 & 2.9 &    &           & $\leq$264 &    &    &    & D \\
2007-Apr-23 & 7.0 &    &           & $\leq$348 &    & $\leq$1421 &    & D \\
2007-Apr-24 & 7.8 &    & $\leq$390 & $\leq$387 &    &    &    & D \\
2007-Apr-25 & 8.8 &    &           & $\leq$417 &     & 1293 $\pm$ 220 &    & D \\
2007-Apr-26 & 9.8 &    & $\leq$339 & $\leq$555 & $\leq$666 &    &    & D \\
2007-Apr-30 & 13.8 &    & $\leq$291 & 480 $\pm$ 102 &    &    &    & D \\
2007-May-03 & 16.8 &    &     &     & $\leq$978 & 1032 $\pm$ 75 &    & D \\
2007-May-05 & 19.2 &    & $\leq$1692 & 753 $\pm$ 107 &    &    &    & D \\
2007-May-12 & 26.1 &    & $\leq$1905 & 804 $\pm$ 100 & 1900 $\pm$ 300 & 1062 $\pm$ 245 &  & D \\
2007-May-17 & 30.9 &    & $\leq$657 & 728 $\pm$ 87 & 1480 $\pm$ 289 & $\leq$1050 &    & D \\
2007-May-27 & 41.3 & $\leq$171 & 487 $\pm$ 96 & 1257 $\pm$ 53 & 1250 $\pm$ 270 & $\leq$1020 & & A\\
2007-Jun-11 & 55.9 &    & $\leq$1191 & 1490 $\pm$ 104 & 2000 $\pm$ 308 & 1246 $\pm$ 127 &    & A \\
2007-Jun-22 & 66.8 &    & 709 $\pm$ 60 & 1390 $\pm$ 45 &    &    &    & A \\
2007-Jul-03 & 77.8 &    &     &     & $\leq$843 & $\leq$978 &    & A \\
2007-Jul-07 & 81.8 &    & 915 $\pm$ 62 & 1325 $\pm$ 47 &    &    &    & A \\
2007-Jul-24 & 98.8 & $\leq$201 & 886 $\pm$ 80 & 1131 $\pm$ 59 & $\leq$909 & $\leq$870 &    & A \\
2007-Aug-18 & 124.0 &    & 1050 $\pm$ 90 & 957 $\pm$ 64 & $\leq$957 & $\leq$693 &    & A \\
2007-Sep-07 & 144.0 & $\leq$336 & 1103 $\pm$ 88 & 621 $\pm$ 81 & $\leq$1137 & $\leq$825 &    & A \\
2007-Sep-23 & 159.8 & $\leq$525 & 792 $\pm$ 77 & 316 $\pm$ 57 &    &    &    & AnB \\
2007-Oct-23 & 189.9 & $\leq$585 & 685 $\pm$ 62 & 379 $\pm$ 59 &    &    &    & AnB \\
2007-Nov-17 & 214.9 & $\leq$342 & 479 $\pm$ 65 & 404 $\pm$ 62 &    &    &    & B \\
2007-Dec-23 & 250.9 & $\leq$387 & 685 $\pm$ 65 & 783 $\pm$ 57 &    &    &    & B \\
2008-Jan-03 & 261.8 &    &     &     &    & 1723 $\pm$ 77 &    & B \\
2008-Jan-05 & 263.8 &    &     &     &    &    & $\leq$1662 & B \\
2008-Jan-28 & 286.8 &    & $\leq$594 & 1669 $\pm$ 90 &  & $\leq$999 &    & B \\
2008-Feb-25 & 314.8 &    & 1059 $\pm$ 70 & 2097 $\pm$ 73 &    & 2570 $\pm$ 160 &    & CnB \\
2008-Apr-19 & 368.8 & $\leq$822 & 1304 $\pm$ 62 & 2200 $\pm$ 59 &    &    &    & C \\
2008-May-07 & 386.8 &    &     & 2852 $\pm$ 58 &    &    &    & C \\
2008-Jun-09 & 419.9 &    & 2039 $\pm$ 88 & 3344 $\pm$ 81 &    & 2023 $\pm$ 140 &    & DnC \\
2008-Nov-03 & 566.9 & $\leq$321 & 2812 $\pm$ 60 & 3897 $\pm$ 78 &    & 1854 $\pm$ 128 &    & A \\
2008-Dec-30 & 623.8 &    &     & 3891 $\pm$ 78 & $\leq$2358 &    &    & A \\
2009-Apr-06 & 720.8 & 746 $\pm$ 168 & 3390 $\pm$ 136 & 3842 $\pm$ 69 &    & 1629 $\pm$ 115 &   & B\\
2009-May-31 & 775.8 & $\leq$1167 & 3460 $\pm$ 80 & 3641 $\pm$ 79 &    &    &    & CnB \\
2009-Aug-27 & 863.8 & $\leq$1503 & 3373 $\pm$ 84 & 3408 $\pm$ 109 &    &    &    & C \\
\hline
\multicolumn{9}{l}{$^{\mathsf{a}}$ Days since explosion assuming an explosion
date of 2007 April 16.15 UT.}\\
\multicolumn{9}{l}{All upper limits are $3\sigma$.}
 \end{tabular}
 \end{minipage}
\end{table*}

\subsection{X-Ray}

SN 2007bg was observed seven times with the {\it Swift} X-ray Telescope (XRT)
and once with the {\it Chandra} X-ray Observatory, as listed in Table \ref{tab:xray}. Processed data
were retrieved from the {\it Swift} and {\it Chandra} archives, and analysis was performed using the
HEASOFT
(v6.12)\footnote{\url{http://heasarc.gsfc.nasa.gov/lheasoft/}} and
CIAO (v4.4.1)\footnote{\url{http://asc.harvard.edu/ciao/}} software packages, along with the latest
calibration files available at the time.  In the case of {\it Chandra}, we reprocessed the data to
apply the latest calibration modifications, apply positional refinements, and correct for charge
transfer inefficiency (CTI).  All datasets were then screened for standard grade selection,
exclusion of bad pixels and columns, and intervals of excessively high background (none was found).

The seven {\it Swift} datasets were combined into three rough epochs.
We adopted a 35\arcsec radius aperture (corresponding to $\sim$85\% encircled energy) as the source
extraction region and modelled the local diffuse and scattered background using an annulus of
35--100\arcsec. No statistically significant detection was found within our
adopted aperture. For the {\it Chandra} dataset, we adopted a 1\farcs5 radius aperture
(corresponding to $\sim$95\% encircled energy) as the source extraction region and modelled the
local diffuse and scattered background using an annulus of 1.5--10\arcsec. Although five (5) counts
fell within our aperture (a $\sim2\sigma$ signal), no statistically significant detection could be
made. Upper limits in all cases were determined using the Bayesian technique from
\citet*{Kraft1991}. We used PIMMS v4.6 to determine a count-to-flux conversion, adopting an absorbed
APEC thermal plasma model with $kT=4$ keV and $N_{\rm{H}}=1.5\times10^{20}$ cm$^{-2}$
\citep{Dickey1990}.

\begin{table*}
\centering
\begin{minipage}{150mm}
\caption{\protect{X-Ray measurements for SN 2007bg} \label{tab:xray}}
\begin{tabular}{lrrcp{0.9cm}cc}
Date$^{\mathsf{a}}$ & Instrument & Obsid  & Exposure & 0.5-8.0 & $F_{0.5-8.0}$ 
& $L_{0.5-8.0}$\\
& & (ks) & & Counts & (erg s$^{-1}$ cm$^{-2}$) & (erg s$^{-1}$)\\
\hline
\hline
2007-Apr-22$^{1}$ & {\it Swift} XRT & 00030920001 & 9.4 & $<$4.6   &
$<2.0\times10^{-14}$ & $<5.5\times10^{40}$\\
2007-Apr-30$^{2}$ & {\it Chandra} ACIS-S3 & 7643 & 34.0 & $<$11.2  &
$<2.6\times10^{-15}$ & $<7.2\times10^{39}$\\
2008-Mar-05$^{3}$ & {\it Swift} XRT & 00030920002 & 2.8 & 
\multirow{3}{*}{$<$11.1} & \multirow{3}{*}{$<4.4\times10^{-14}$} &
\multirow{3}{*}{$<1.2\times10^{40}$} \\
2008-Mar-17$^{3}$ & {\it Swift} XRT & 00030920003 & 4.2 & & &\\
2008-Apr-23$^{3}$ & {\it Swift} XRT & 00030920004 & 2.9 & & &\\
2009-Jun-09$^{4}$ & {\it Swift} XRT & 00030920005 & 3.5 & 
\multirow{3}{*}{$<$6.9} & \multirow{3}{*}{$<3.1\times10^{-14}$} &
\multirow{3}{*}{$<8.6\times10^{40}$} \\
2009-Jun-15$^{4}$ & {\it Swift} XRT & 00030920006 & 3.1 & & &\\
2009-Jun-17$^{4}$ & {\it Swift} XRT & 00030920007 & 2.0 & & &\\
\hline
\multicolumn{7}{l}{$^{\mathsf{a}}$ Different epochs were combined to increase SNR. The
superscript on the date indicates which were combined.}\\
\end{tabular}
\end{minipage}
\end{table*}

\begin{figure*}
 \includegraphics[width=17cm]{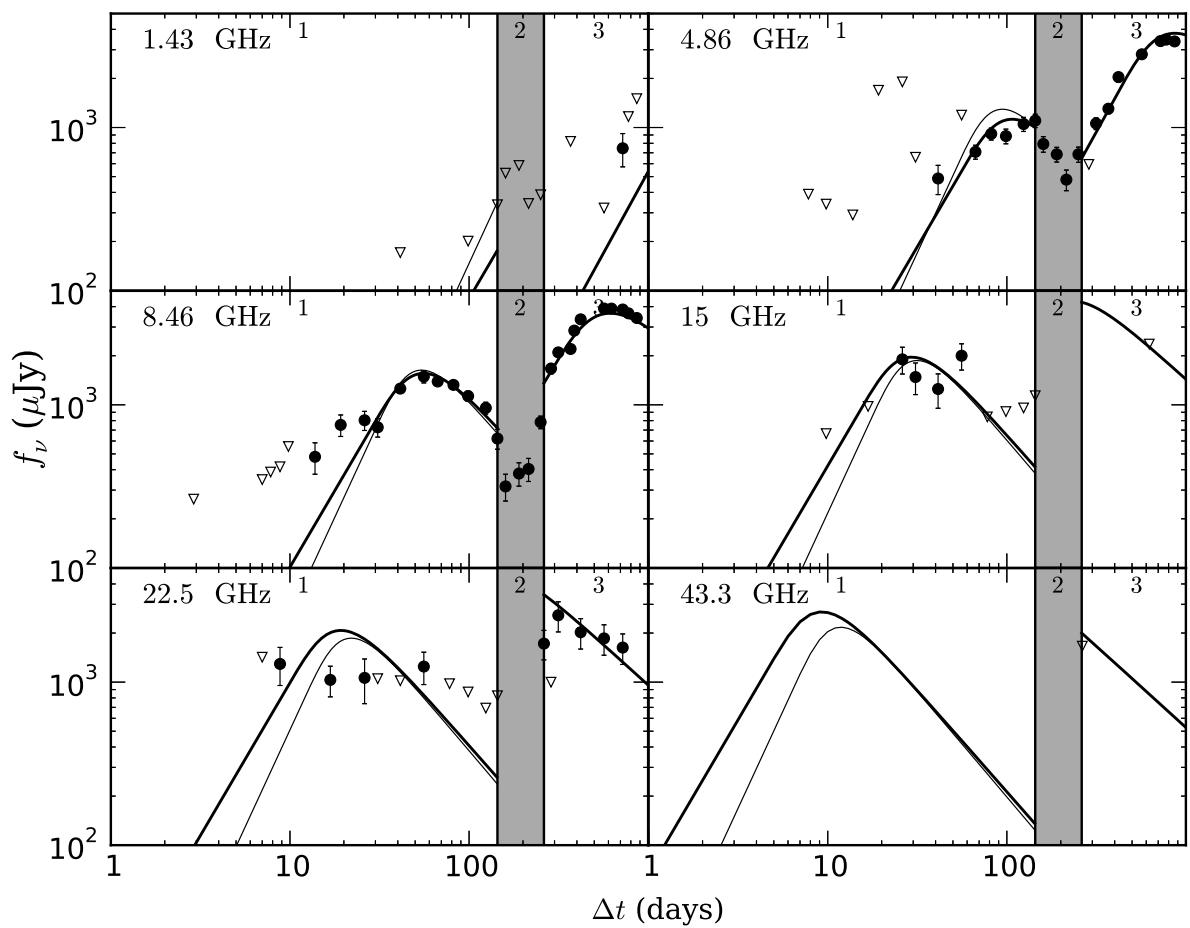}
  \caption{\label{f:lc} Light curves for SN 2007bg. Phase 1 is shown as the first white region,
phase 2 is the shaded region and phase 3 the last white region. Different frequencies are
represented by the following colours: $1.43$ GHz, $4.86$ GHz, $8.46$ GHz, $15$ GHz, $22.5$ GHz and
$43.3$ GHz. We see a slow turn-on at $8.46$ GHz during phase 1.
The lines during phase 1 represents the model fits. The thin line assumes a density profile
$r\propto t^{0.94}$ and $B\propto t^{-1}$, while the thick line requires $r\propto t^{0.75}$ and
$B\propto t^{-0.75}$. During this phase we modelled the emission as arising from a spherical
expanding shell self that is absorbed by its own radiating electrons, as explained in 
\S\ref{ssec:ssa}. During phase 3 the line model represents our best fit model including the
presence of SSA and a clumpy CSM. This is described in \S\ref{ssec:phase3}. Filled circles denote
detections with $1\sigma$ error bars (which are smaller than symbols in some cases), while inverted
triangles denote $3\sigma$ error limits.}
\end{figure*}

\begin{figure*}
\includegraphics[width=15cm]{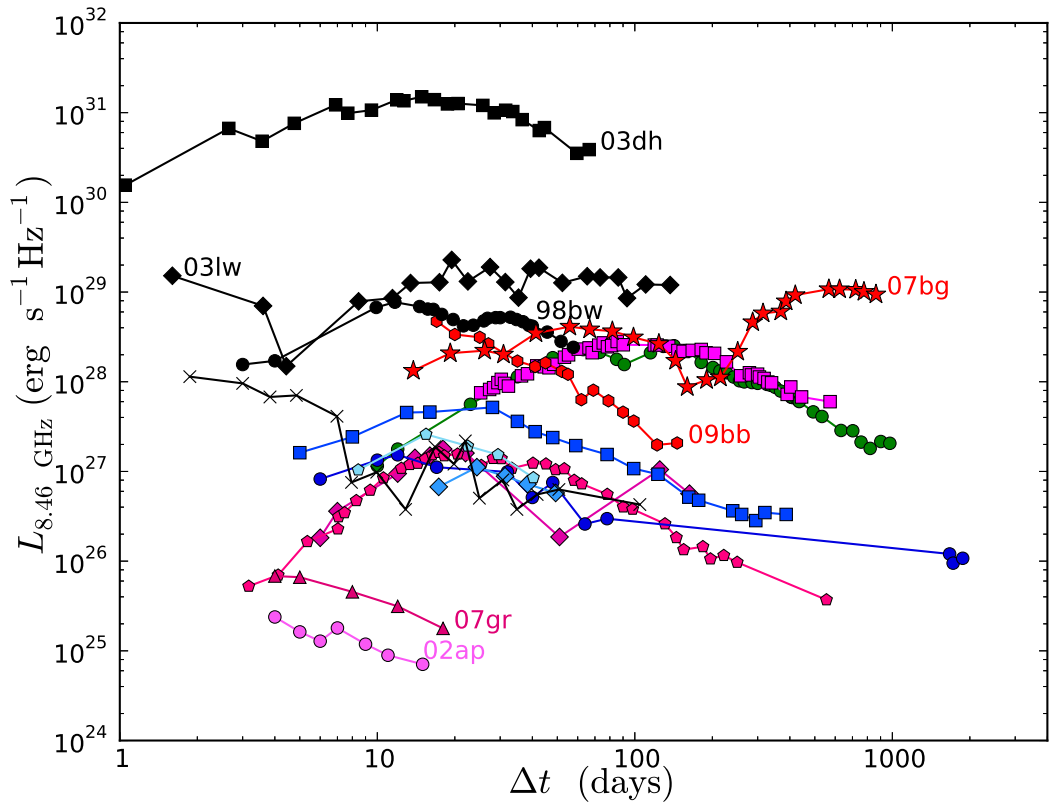}
  \caption{\label{f:msne} Radio light curves for different Type Ib SNe (shades of blue with
different symbols) Type Ic's (shades of pink with different symbols), a Type IIb SN (green) and GRBs
(black with different symbols). SN 2007bg (red stars) is shown along with other objects. We observe
that the radio luminosity from SN 2007bg is lower than that of GRBs on comparable time-scales and it
is larger than in regular SNe at the time of the first peak during phase 1. Phase 3 starts at a
comparable time-scale compared to other SNe, but the peak luminosity reached is larger by orders of
magnitude. The SNe shown in this plot are Type Ic 2002ap \citep*[circles;][]{2002ApJ...577L...5B},
Type IIb 2003bg \citep[circles;][]{2006ApJ...651.1005S}, Ic 2003L
\citep[squares;][]{2005ApJ...621..908S}, Ic 2004cc, Ib 2004dk, Ib 2004gq \citep[diamonds; circles;
squares;][]{Wellons2012}, Ic 1994I \citep[pentagons;][]{2011ApJ...740...79W}, Ic 2007gr
\citep[triangles;][]{2010ApJ...725..922S}, Ib 2007uy, Ib/X-ray flash 2008D \citep[diamonds;
triangles;][]{2011ApJ...726...99V} and Type Ic broad-lined 2009bb
\citep[hexagons;][]{2010Natur.463..513S}. Also, SNe associated with a GRBs Type Ic 1998bw
\citep[circles;][]{1998Natur.395..663K}, Ic 2003dh 
\citep[squares;][]{Hjorth2003,Kawabata2003,Matheson2003,2005ApJ...619..994F}, Ic 2003lw
\citep[diamonds;][]{2004Natur.430..648S} and Ic 2006aj \citep[Xs;][]{2006Natur.442.1014S}.}
\end{figure*}

\begin{figure*}
  \includegraphics[width=15cm]{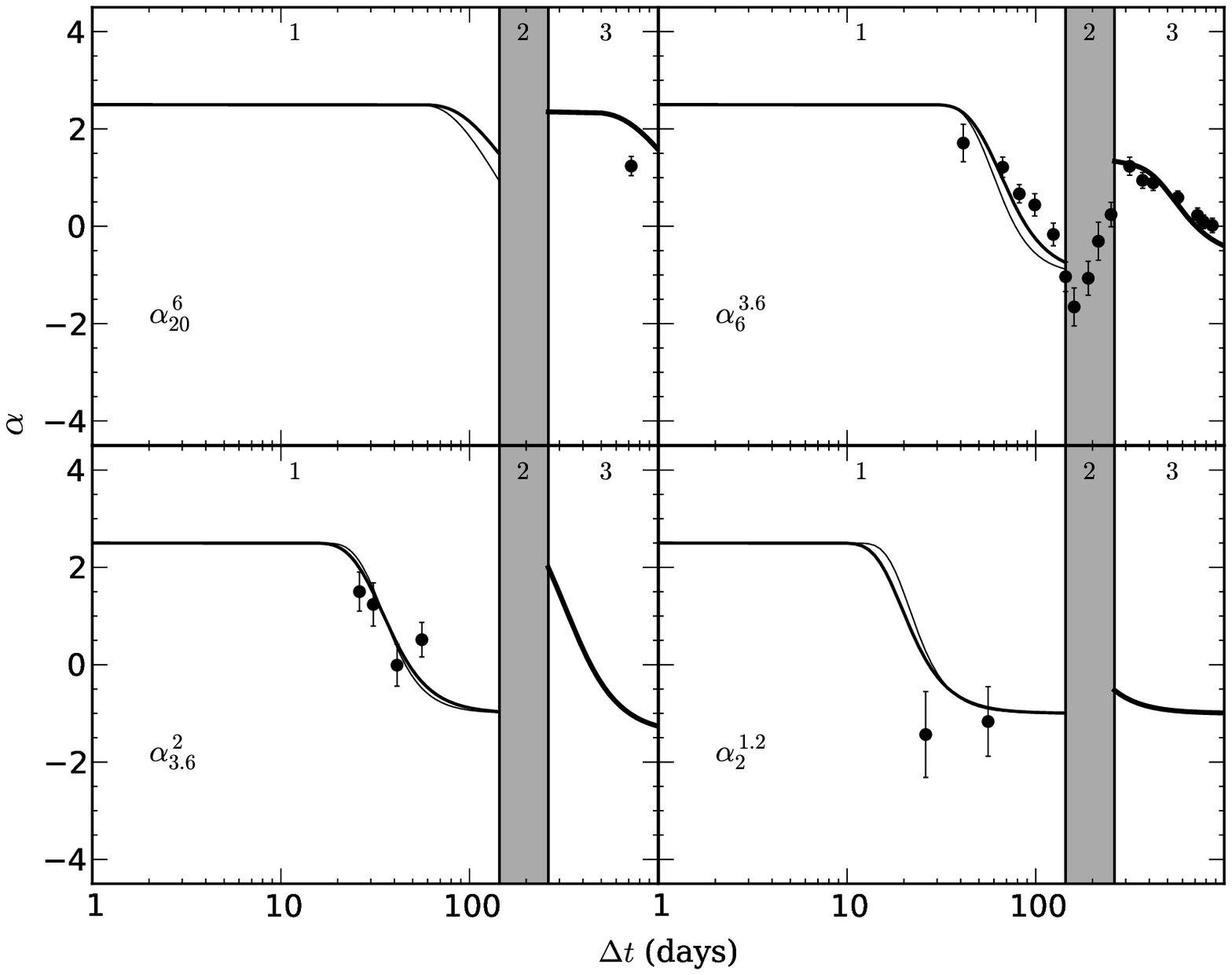}
  \caption{\label{f:si} Spectral indices between adjacent bands. Over plotted are
the predicted values from the light curve fitting as described in \S\ref{ssec:ssa} and
\S\ref{ssec:phase3}. The thin and thick lines represent the same models as in Fig. \ref{f:lc}.
{\it Top left} panel shows the evolution of the spectral index between $20$ cm
and $6$ cm. {\it Top right} panel shows the spectral index from $6$ cm and
$3.6$ cm, the best sampled set of observations. {\it Bottom left} spectral
index between $3.6$ cm and $2$ cm. {\it Bottom right} spectral index between $2$ cm and $1.2$ cm. We
note that near the end of phase 1 a large negative spectral index is observed, this is mainly
because the light curve at $1.2$ cm deviates from the evolution observed at other wavelengths.
Filled circles denote detections with $1\sigma$ error bars (which are smaller than symbols in some
cases), while inverted triangles denote $3\sigma$ upper limits and triangles $3\sigma$ lower
limits.}
\end{figure*}

\section{Results}
\label{s:results}

The resulting light curves for SN 2007bg are shown in Fig. \ref{f:lc}. The evolution of the light
curves can be divided in three phases. \textit{Phase 1}: this phase starts after the explosion and
lasts until day $144$.
Here we observe the typical turn-on of radio supernovae (RSNe), first starting at shorter
wavelengths and cascading to longer wavelengths with time as wavelength-dependent absorption becomes
less important. The peak luminosity reached during phase 1 is $4.1\times10^{28}$ \ergphzps\ observed
on day $55.9$ at $8.46$ GHz. This makes SN 2007bg one of the most luminous Type Ib/c RSNe on
comparable time-scales. The first detections at $4.86$ and $8.46$ deviate from the rest of the
evolution, and it could be due to interstellar scintillation, a real deviation from pure SSA due to
clumpy CSM, or deceleration; unfortunately there is not enough early-time data to confirm it
robustly. On day $55.9$ we also observe a rise in luminosity at $15$ and $22.5$ GHz, this could be
the effect of a small density enhancement, since that would produce the achromatic variation
observed, and as the lower frequencies are still optically thick it would not be noticed at these
frequencies. \textit{Phase 2}: if the shock is expanding into a constant (or even smoothly-varying)
stellar wind, then after the absorption turn-on
one expects the source to fade in intensity following a power-law decline with time. However on day
$144$ the flux density consistently deviates from this at $8.46$ and $4.86$ GHz, showing a
discontinuous drop in the observed flux density from days $\sim144-250$. 
\textit{Phase 3}: the luminosity of SN 2007bg begins a second turn-on around day $261.8$, becoming
brighter than in phase 1 and reaching a peak luminosity of $1\times10^{29}$ \ergphzps\ at $8.46$ GHz
on day $566.9$. This late time behaviour is comparable to that of other Type
Ib/c SNe like SN 2004cc, SN 2004dk and SN2004gq \citep*{Wellons2012}, albeit much stronger and more
well-defined than these other objects. There are also some Type IIb's SNe which show late time
modulations in their radio light curves. One of these is SN 2001ig \citep{2004MNRAS.349.1093R} in
which case the late time modulations could be explained by a binary companion of the progenitor.
Another example is SN 2003bg \citep{2006ApJ...651.1005S}, where the late time variations can not
rule out a binary companion or different mass-loss episodes of the progenitor. In both cases the
modulations in the light curve result from modulations in the CSM density.
A comparison between SN 2007bg, other SNe and GRBs is shown in Fig. \ref{f:msne}. During phase 1 the
light curve of SN 2007bg resembles that of other SNe. It shows the characteristic turn-on, but with
a higher luminosity than other normal Ib/c SNe and comparable to GRB-SNe. Then during phase 3 we
observe a strong rise in its luminosity, which is unprecedented for SNe, as it reaches such high
luminosities, $2.6$ times larger than during phase 1.

The spectral index, $f_{\nu}\propto\nu^{\alpha}$, evolution of the source is shown in Fig.
\ref{f:si}.
We observe a trend towards a negative value between the best sampled frequencies $4.86$ GHz and
$8.46$ GHz during phase 1, but it could be that during this phase the source is still optically
thick at $4.86$ GHz. If we assume that at this time the emission has reached its peak luminosity at
$4.86$ GHz, then the late observed values of the spectral index point to a value of
$\alpha\approx-1$. Even the deviant points at $15$ and $22.5$ GHz on day $55.9$ are consistent with
this spectral index value. During phase 2 the source spectrum begins to invert. This
behaviour continues during the first days of phase 3, indicating that some absorption mechanism
becomes important again in phase 3.
From the earliest spectral indices available we observe that the values are $\alpha\sim2$, which is
lower than the typical synchrotron self absorption (SSA) value of $2.5$ and the spectrum is
shallower than the exponential cut-off expected from free-free absorption (FFA). This suggests may
be observing a variety of emission regions from SN 2007bg; some being absorbed
more highly than others. Since we can not resolve the source, we get the combined flux resulting in
a shallower optically thick part of the spectrum.

Given that we do not have strong constraints on how the intrinsic spectral index varies with time,
we adopt a fixed spectral slope for each phase of $\alpha\approx-1$.

\subsection{Early Physical Constraints from Phase 1}
\label{ssec:early_ssa}

\begin{figure*}
\includegraphics[width=150mm]{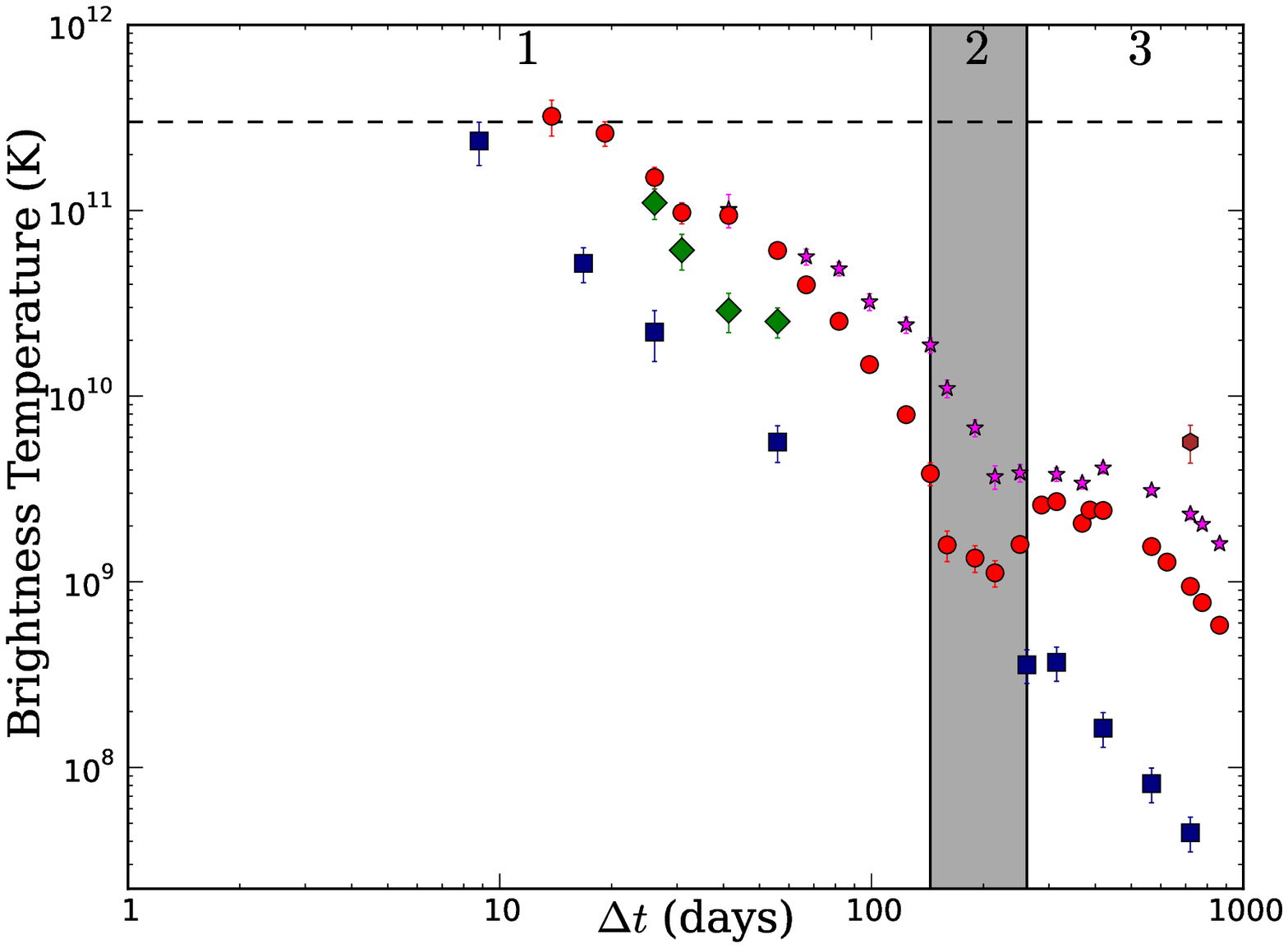}
  \caption{\label{f:tb} Brightness temperature curve for SN 2007bg using a
constant bulk speed of $0.18c$. Different wavelengths are represented by different symbols, $1.43$
GHz (brown hexagons), $4.86$ GHz (magenta stars), $8.46$ GHz (red circles), $15$ GHz (green
diamonds) and $22.5$ GHz (blue squares).
The dashed line represents the ICC limit of $3\times10^{11}$ K. The symbols are shown with
their $1\sigma$ error bars (which are smaller than symbols in some cases). $3\sigma$ upper limits
are represented by inverted triangles following the colour coding.}
\end{figure*}

Our first data point at $8.46$ GHz allows us to evaluate the brightness temperature of the source at
$13.8$ days after explosion. The brightness temperature of the source should not exceed
$3\times10^{11}$ K if it is to remain under inverse Compton catastrophe (ICC), where the cooling
time due to inverse Compton scattering is shorter than the synchrotron cooling time
\citep*{1994ApJ...426...51R} \citep*[see also;][]{1969ApJ...155L..71K}. Using the measured flux
density of $753$ \mJy\ at $8.46$ GHz, and if we assume that the optical expansion speed of
$\sim17500$ km s$^{-1}$, measured from the optical Ca II triplet, at $\sim5$ days after explosion
\citep{2010A&A...512A..70Y} also pertains the radio emitting region, we
obtain $T\approx3.6\times10^{12}$ K.
This value is above the ICC limit, so one of our assumptions must be wrong. As discussed in the
introduction, typically the radio ejecta from SNe have higher velocities than that of the
photosphere, which produces the optical emission. Based on this we adopt a radius for the
source such that the brightness temperature remains below the ICC limit constraining the source size
to be $(7\pm1)\times10^{15}$ cm. The reported errors come from the flux density measurement
uncertainty. At $13.8$ days after explosion this implies a bulk expansion speed
of $\beta\Gamma\approx(0.2\pm0.04)c$, where $\beta=v/c$ and $\Gamma$ is the Lorentz factor. This
speed $v\approx(0.19\pm0.03)c$ is $\approx3$ times larger than that of the photosphere. We stress
that this only a lower limit on the actual velocity at which the radio emitting region expands,
as the radio ejecta could be expanding at a higher velocity and be well below the ICC brightness
temperature.
The resulting brightness temperature is shown in Fig. \ref{f:tb} where we assumed that the radio
emitting region does not decelerate i.e., $r\propto t$.

\begin{figure*}
\includegraphics[width=17cm]{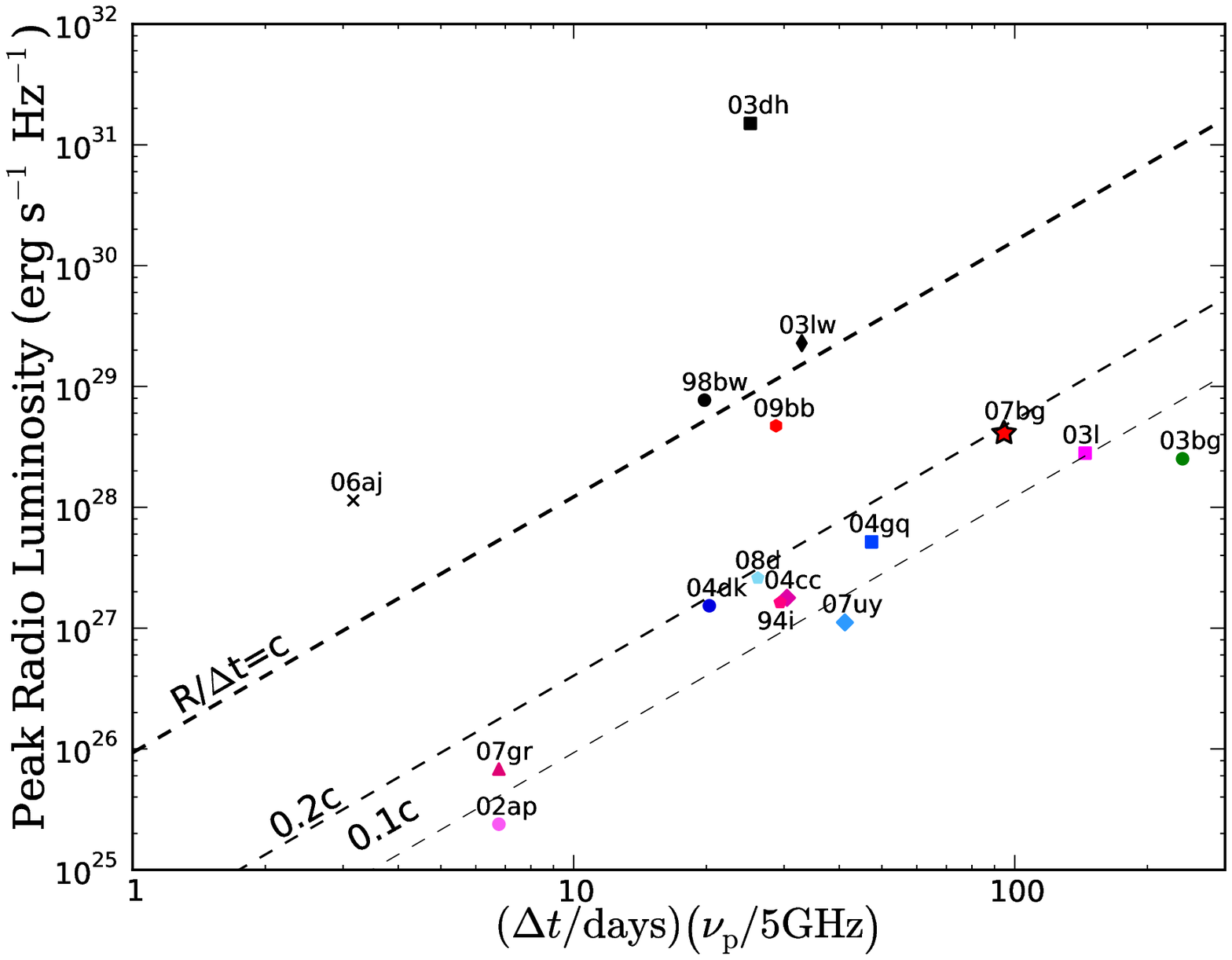}
  \caption{\label{f:vel} Source peak radio luminosity versus time of peak. The resulting position in
this plot provides an estimate of the average expansion speed of the source. Some objects lie above
$R/\Delta t=c$, because the emission from GRBs is different from that of SNe, for which this
diagnostic was made. SN 2007bg lies at the high end of normal Type Ib/c SNe based on its high
luminosity and shows an expansion speed of $\gamma\beta\approx0.2c$. Here we have used data from
phase 1 only.
Other SNe and GRBs are the same as in Fig. \ref{f:msne}. Error bars are small compared to the
plot symbols.}
\end{figure*}

Another way to estimate the expansion speed of the radio emitting region can be obtained by placing
SN 2007bg in a luminosity peak time plot, which as discussed by
\citet{1998ApJ...499..810C} provides information about the average expansion speed of the radio
emitting region. The plot is shown in Fig. \ref{f:vel} and we note that SN 2007bg has an average
expansion speed of $\sim0.2c$ typical of Type Ib/c RSNe and it shows one of the highest luminosities
among Type Ib/c's and comparable to those of GRBs. It is important to note that  other absorption
mechanisms, such as FFA, or deviations from equipartition, would increase the expansion velocity
(and hence source energy, derived later).

\subsection{Constraints from SSA Evolution during Phase 1}
\label{ssec:ssa}

\noindent
Synchrotron emission from SNe originates from the shock region between the SNe
ejecta and the pre-existing CSM. This shock amplifies the existing magnetic fields via
turbulence. The electrons in the shock region then interact with this magnetic field producing the
observed synchrotron emission. This emission is subject to self-absorption, which is caused by SSA
\citep{1998ApJ...499..810C} in the case of low density environments.
To describe the time evolution of the flux density during phase 1 we adopt the parametrization of
\citet{2005ApJ...621..908S} in which the radio emission originates from a spherical thin shell of
width $r/\eta$ expanding at a velocity $v$ in which the intrinsic synchrotron emission is absorbed
by SSA. The parameter $\eta$ is the number of shells that fit into the size of the source with a
value of $\eta\approx10$ \citep*{2000ApJ...537..191F}. 
We adopt the assumptions of the standard model of \citet{1982ApJ...259..302C} where the hydrodynamic
evolution of the ejecta is self-similar across the shock discontinuity. With this assumption the
radius of the source evolves as $r\propto t^{\alpha_{r}}$ where the parameter $\alpha_{r}$
is related to the density profile of the outer SN ejecta $\rho\propto r^{-n}$ and the density
profile of the radiating electrons in the shock region $n_{\rm{e}}\propto r^{-s}$ such that
$\alpha_{r}=(n-3)/(n-s)$. The self similar model also requires that $s<3$, $n>5$
and $\alpha_{r}$ in the range $0.75-1$. The model also assumes that the magnetic energy density and
the relativistic electron energy density scale as the total post-shock energy density. 
This implies that the fraction of energy contained in the relativistic electrons
$\epsilon_{\rm{e}}$ and magnetic field $\epsilon_{\rm{B}}$ remains fixed. This
in turn relates the magnetic field evolution $B\propto t^{\alpha_{B}}$, to the radius of the source
and the density profile of the radiating electrons $\alpha_{B}=(2-s)\alpha_{r}/2-1$. This means that
for a freely expanding wind $s=2$, the magnetic field evolves as $B\propto t^{-1}$.
For our analysis we adopt a source in equipartition with $\epsilon_{\rm{e}}=\epsilon_{\rm{B}}=0.1$.
Equipartition of energy minimizes the total energy.

Using this model and requiring the time evolution of the magnetic field to go as $B\propto t^{-1}$
we find that the radius of the source evolves as $r\propto t^{0.94}$. We will refer to this as
model 1. This is consistent with a shock expanding freely through the CSM. The normalization
constants for the flux density and SSA are $C_{f}\approx8.4\times10^{-53}$ g s$^{-1}$ and
$C_{\tau}\approx8.5\times10^{36}$ s$^{3.5}$, giving a reduced chi square value of $\chi_{r}=7.3$ for
22 degrees of freedom. These values require the size of the source to go as
$r\approx6.5\times10^{15}(t/t_{0})^{0.94}$ cm, which in turn means
that the expansion velocity is $v\approx0.25(t/t_{0})^{-0.06}c$. The magnetic field  goes as
$B\approx2.6(t/t_{0})^{-1}$ G and the energy is $E\approx9.6\times10^{47}$ erg. For this we assumed
a synchrotron characteristic frequency of $\nu_{\rm{m}}=1$ GHz. Using these values we can also
estimate the mass-loss rate of the progenitor. Following \citet{2006ApJ...651.1005S} we estimate a
mass-loss rate of $\dot{M}\approx1.5\times10^{-6}(v_{w}/1000\;\rm{km\,s}^{-1})$ \Msunpyr. This
mass-loss rate lies at the low end of those observed for WR stars in low metallicity environments
\citep{Crowther2002}. The derived mass-loss rates from this model depend on the
fraction of energy contained in relativistic electrons and magnetic fields, deviations from
equipartition would require higher mass-loss rates and energy requirements.

The early data hint at a shallower slope of the light curve, so we also fit a second model, model 2,
where we allow the magnetic field evolution to change as $B\propto t^{-0.75}$
and the source size as $r\propto t^{0.75}$. In this model the reduced chi square value is
$\chi_{r}=4.4$ for 21 degrees of freedom. In this case the normalization constants are
$C_{f}\approx1.7\times10^{-52}$ g s$^{-1}$ and $C_{\tau}\approx4.6\times10^{36}$ s$^{3.5}$. This
implies a source size evolution of $r\approx8.7\times10^{15}(t/t_{0})^{0.74}$ cm, and an expansion
velocity of $v\approx0.3(t/t_{0})^{-0.25}c$. Given this expansion speed the source would remain
below the ICC limit. 
The magnetic field goes as $B\approx2.1(t/t_{0})^{-0.75}$ G. The source energy with these
values is $E\approx1.5\times10^{48}$ erg, and the mass-loss rate is
$\dot{M}\approx1.9\times10^{-6}(v_{w}/1000\;\rm{km\,s}^{-1})$ \Msunpyr. Here
again we have used the same parameters for $\nu_{m}$ and $\eta$. Given this time evolution for the
magnetic field and source size, the density radial profile goes as $\rho\propto r^{-8}$ and the
number density of emitting electrons as $n_{\rm{e}}\propto r^{-1.3}$.
We adopt the result from model 2, as it provides a better fit to the data, although we cannot truly
rule out model 1 due to the lack of early detections.

Given this mass-loss rate, the circumstellar density for a wind stratified medium $A_{*}$, defined
as $\dot{M}/4\pi v_{w}=5\times10^{11}A_{*}$ g cm$^{-1}$ \citep*{Li1999}, is $A_{*}\approx0.2$. Under
these conditions a relativistic jet from a GRB would become non-relativistic, and hence observable
on a time-scale of $t_{\rm{NR}}\approx0.3(E_{\rm{K}}/10^{51}\,\rm{erg})A_{*}^{-1}$. With the
kinetic energy of the ejecta of $E_{\rm{K}}\approx4\times10^{51}$ erg, derived from the optical
light curve \citep{2010A&A...512A..70Y}, this time-scale is $t_{\rm{NR}}\approx6$ yr. The lack of
clear afterglow signatures \citep[e.g.,][]{Li1999,Waxman2004,2006ApJ...638..930S} in our
observations thus provide strong evidence against an expanding GRB jet being the source of the
observed luminosity for SN 2007bg.

\subsection{Stratified CSM: Phase 2}

As can be seen in Fig. \ref{f:lc}, at around day $144$ the flux density at $8.46$ and $4.86$ GHz
suddenly decreases creating a break in the power law flux decay of the optically thin radio
emission. At $8.46$ GHz the observed flux is a $44\%$ less than the predicted one at the beginning
of phase 2 on day $159.8$.
Based on the derived size of the radio emitting region, from the SSA modelling, this drop occurs at
a distance of $\approx6.7\times10^{16}$ cm.
The most likely cause for this decrease in the radio emission is a drop in the CSM density. Here we
consider two scenarios that would produce a change in CSM density such as a binary companion or two
different wind components from the progenitor star. In the first scenario, there should be evidence
of this interaction from the optical spectra as Balmer series recombination lines. The late time
optical spectra for SN 2007bg was too faint to be detected. The early optical
spectra evidence points to a broad-lined SNe, most likely associated with SNe like 1998bw
\citep{2010A&A...512A..70Y}, and the latest optical spectra of SN 2007bg where it is detected does
not show evidence for a transition from a type Ib/c to a type II. For this kind of SNe the most
likely progenitor is a single WR. Also if there was a binary companion we would expect to see
coherence between the emission before and after the drop/increase (depending on whether the CSM
density dropped/increased at the region of interest). In the case of SN 2007bg, we see a $>1$ dex
jump in the normalizations of the unabsorbed phase 1 and phase 3 light curves, as well as a strong
enhancement in absorption, making a binary companion CSM scenario highly unlikely. The second
scenario considered is that the drop in flux density is due to a change in the progenitor wind.
Based on the distance of the drop and with our assumed wind speed of $\approx1000$ km s$^{-1}$ the
change occurred $\approx45$ yr before the explosion and lasted for $\approx23$ yr.
Since $L_{\nu}\propto(\dot{M}/v_{w})^{(\alpha-8+12\alpha_{r})/4}$
\citep{1982ApJ...259..302C} a flux density decrease of $44\%$ implies that the
mass-loss rate/wind speed dropped by a factor of $\approx5$. For this estimate we have used
$\alpha_{r}=0.94$. Our modelling of the light curves during phase 1 suggests that the density
profile of the CSM is shallower (i.e., smaller $\alpha_{r}$), but as discussed bellow, this
relation overestimates the mass-loss rate in such cases.

\subsection{Second Rise: Phase 3}
\label{ssec:phase3}

Phase 3 begins around day $261.8$ with an increase in the flux density at $22.5$,
$8.46$ and $4.86$ GHz. As discussed in \S\ref{s:results} the luminosity of the source during this
phase is larger than in phase 1 in all of the bands where we detect it. This flux density
rise shows the characteristics of an absorption turn-on. At $22.5$ GHz we observe what appears to be
the peak of the turn-on on day $314.8$, and at longer wavelengths the radio spectrum begins to
turn-on near the end of observations. This turn-on is accompanied by an increase in the spectral
index suggesting that the source has become optically thick again. 
The luminosity increase implies a sharp CSM density enhancement, presumably due to evolution of the
progenitor wind, while the observed re-absorption requires either SSA or FFA. Following
\citet*{Fransson1996} we can estimate the mass-loss rate from the progenitor assuming that the
absorption is purely FFA. Adopting the time of peak at $4.86$ GHz as the time when the free-free
optical depth is unity, and taking the expansion velocity of the shock from phase 1 then the
mass-loss rate is $\dot{M}\approx10^{-1}(v_{w}/1000\;\rm{km\,s}^{-1})$ \Msunpyr. This rate is
suspiciously high and inconsistent with our estimate (later in this section) of the mass-loss rate
based on the unabsorbed portion of the spectrum. This high rate favours SSA. The absorption slope
is relatively flat, and thus pure SSA with a similar model as fit to phase 1 does a poor job of
fitting the turn-on. One possibility is differential SSA arising from small dense clumps in the CSM,
which are preferentially affected by SSA while sparser regions are less affected; this naturally
flattens the optically thick slope. The existence of clumpy winds in WR stars has been confirmed by
observations \citep{1988ApJ...334.1038M}, and thus provides a natural way of explaining the observed
radio emission. Alternatively, the density jump would naturally produce strong X-ray emission, which
could ionize CSM material in front of the shock and provide the necessary free electrons for FFA.
However, without constraints on the X-ray luminosity and phase 3 CSM density distribution for
SN2007bg, it is difficult to assess the feasibility of FFA.
Given the poor fit of the pure SSA model ($\chi_{r}=9$), we simply adopt the parametrization of
\citet{2002ARA&A..40..387W} including only a clumpy CSM with the presence of internal SSA. In this
case the flux density is normalised by $K_{1}$, the clumpy CSM absorption by $K_{3}$ and SSA by
$K_{5}$.
We fit the above model to the light curve during phase 3. The best-fit values
are given in Table \ref{t:phase3}. The resulting reduced chi square of this fit is $\chi_{r}=4.4$
for 18 degrees of freedom.

\begin{table}
 \centering
 \begin{minipage}{84mm}
 \caption{\protect{SN 2007bg best-fit values for phase 3.}
\label{t:phase3}}
 \begin{tabular}{lc}
  \hline
  \hline
  Parameter & Adopted value \\
  \hline
  $\alpha$      & $-1^{a}$\\
  $\beta$       & $-1^{b}$\\
  $\alpha_{r}$  & $1^{c}$\\
  \hline
  Parameter & Best-fit value\\
  $K_{1}$       & $5\times10^{6}$\\
  $K_{3}$       & $2.3\times10^{5}$\\
  $\delta'$     & $-3.53$\\
  $K_{5}$       & $4.5\times10^{8}$\\
  $\delta''$    & $-2.96$\\
  \hline
  \multicolumn{2}{l}{$^{a}$ Adopted from observed values of $\alpha$.}\\
  \multicolumn{2}{l}{$^{b}$ Assumed because of the lack of data.}\\
  \multicolumn{2}{l}{$^{c}$ Computed using $\alpha_{r}=-(\alpha-\beta-3)/3$.}\\
\end{tabular}
\end{minipage}
\end{table}

As shown in Fig. \ref{f:lc}, this model provides a reasonable fit to the light curves at all
frequencies and also to the spectral index, Fig. \ref{f:si}. The deviant point at $1.43$ GHz
could be easily affected by the host galaxy luminosity.
The data did not allow us to constrain the deceleration rate, so we adopted a constant expansion
by fixing $\beta=-1$, which in turn requires that $\alpha_{r}=m=1$ given the spectral index.
Using our best-fit values we can estimate the mass-loss rate from the progenitor during this
period following equations (11) and (13) of \citet{2002ARA&A..40..387W}. The effective optical depth
at the time of our latest observation, $\sim863$ days, at $4.86$ GHz is
$\langle\tau_{\rm{eff}}^{0.5}\rangle\approx2.2\times10^{-3}$ and $m\approx1$. If we extrapolate the
velocity derived in \S\ref{ssec:ssa} to phase 3, $v_{i}\approx0.13c$ at $t_{i}\approx300$
days, then the derived mass-loss rate is
$\dot{M}\approx4.3\times10^{-4}(v_{w}/1000\;\rm{km\,s}^{-1})$ \Msunpyr, where an electron
temperature of $10^{4}$ K was used. 
This mass-loss rate is higher than the one observed during phase 1, which is a natural explanation
for the high luminosity observed during phase 3. However, such a large mass-loss rate has not yet
been directly observed, and it is at odds with the metallicity mass-loss rate relation derived from
the Milky Way and Magellanic Clouds WR samples \citep[see for e.g.,][]{Crowther2006}. This relation
predicts orders of magnitude lower $\dot{M}$ in such a low metallicity host. On the other hand, this
large mass-loss rate is also roughly consistent with that of the saturation limit for line-driven
winds, $\dot{M}\approx10^{-4}$ \Msunpyr, and is of the same order of magnitude as values found in
other Type Ib/c SNe \citep[e.g.,][]{Wellons2012}.
We should caution however that the dramatic flux increase between phases 1 and 3 implies a strong
density contrast that is inconsistent with one of the primary assumptions of the
\citet{1982ApJ...259..302C} self-similar model (namely a CSM density profile
with a power-law slope). This density discontinuity could naturally lead to an overestimate of the
mass-loss rate; to properly account for this type of case, one should generally model the
hydrodynamics in detail; unfortunately, such simulations are not well-constrained by radio data
alone due to ambiguities in how the B-field and electron density distribution might change across
the density discontinuity and are beyond the scope of this work. Thus this phase 3 mass-loss rate
should be regarded with caution and viewed as a likely upper limit.
The strong mass-loss variations on such short time-scales seen here imply that
the progenitor of SN 2007bg may have undergone unstable mass-loss just prior to explosion or a
dramatic change in wind properties. Such rapid variations have been seen in a handful of objects
before. For instance, in the case of Type Ib SN 2006jc \citep{Pastorello2007}, a bright optical
transient was observed two years prior to explosion. This was interpreted as a luminous blue
variable-like outburst event from the progenitor, which the later SN explosion surely interacted
with. Then there are other possibilities like the case of Type IIn SN 1996cr
\citep{2008ApJ...688.1210B}. In this event, the radio light curve also exhibits a strong late-time
rise in emission associated with a density enhancement likely arising from the interaction of the SN
blast-wave with a wind-blown bubble \citep*{Dwarkadas2010}. The progenitor of SN 1996cr is proposed
to be either a WR or blue supergiant star. We still know relatively little about why such global
changes in the progenitors occur, but it seems clear that a subset of SNe progenitors have them.

\subsection{X-Ray Constraints}

X-ray emission from supernovae can arise from three mechanisms; non-thermal
emission from synchrotron radiating electrons, thermal bremsstrahlung emission from the material in
the circumstellar and reverse shocks, and inverse Compton emission. Since X-ray emission from SN
2007bg could not be detected it is difficult to assess the feasibility of each of these mechanisms,
so we will compare the derived mass-loss rates from the previous subsections with that predicted by
the thermal bremsstrahlung process.

If the X-ray emission comes from synchrotron radiating electrons then the X-ray luminosity can be
obtained by extrapolating the radio synchrotron spectrum up to higher frequencies. To do this we
must consider that the synchrotron spectrum changes its slope at the cooling frequency
$\nu_{\rm{c}}\propto t^{-2}B^{-3}$ by a factor $\Delta\alpha\approx-0.5$
\citep{2005ApJ...621..908S}. Using the derived magnetic field $B$ from our SSA analysis of the
radio light curve \S\ref{ssec:ssa} we find that $\nu_{\rm{c}}\approx323$ GHz. Extrapolation
of the luminosity as $L_{\nu}\propto\nu^{-1.5}$ from the cooling break on gives a X-ray luminosity 5
orders of magnitude bellow our upper limit $\sim30$ days after explosion (epoch 2). Extrapolation
of the cooling frequency up to epochs 3 and 4 of the X-ray data yields X-ray fluxes $\sim5$ orders
of magnitude bellow the upper limits. However, as mentioned in previous sections, the CSM
discontinuity between phases 1 and 3 could cause a break in the magnetic field evolution.

X-ray emission from thermal bremsstrahlung comes from the interaction of the
shocked electrons with the circumstellar material. This interaction heats the electrons in the
medium which then cool via free-free emission. In this process the X-ray luminosity is related
to the density of the emitting material as $L\propto\dot{M}^{2}v_{w}^{-2}t^{-1}$
\citep{Chevalier2003,Sutaria2003}.
We can obtain an upper limit on the mass-loss rate from the X-ray luminosity upper limits using
equation (30) of \citet{Chevalier2006}.
At $t\sim30$ days the X-ray luminosity is $L_{0.5-8.0}<7.2\times10^{39}$ \ergps. We use the density
profile of the outer SNe ejecta derived from our SSA analysis, $\rho\propto r^{-8}$. Given this
luminosity and CSM properties thermal bremsstrahlung requires a mass-loss rate of
$\dot{M}<4.6\times10^{-4}(v_{w}/1000\;\rm{km\,s}^{-1})$ \Msunpyr. At epochs 3 and 4 of the X-ray
data we obtain mass-loss rates of $\dot{M}<3.6\times10^{-3}(v_{w}/1000\;\rm{km\,s}^{-1})$ and
$\dot{M}<1.4\times10^{-2}(v_{w}/1000\;\rm{km\,s}^{-1})$ respectively. These mass-loss rates are
consistent with the one required by SSA during phase 1 and the light curve during phase 3.

Inverse Compton scattering of the optical photons by the relativistic
synchrotron emitting electrons can also be a cause of X-ray emission in SNe. It has been shown that
this process played an important role in the case of SN 2002ap \citep{Bjornsson2004}, SN 2003L
\citep{2005ApJ...621..908S} and 2003bg \citep{2006ApJ...651.1005S}. Using equation (15) of
\citet{Bjornsson2004} we obtain a ratio of $U_{\rm{ph}}/U_{B}\sim12$ $\sim30$ days after explosion.
For this we have used a bolometric luminosity of $L_{bol}\sim10^{41.5}$ erg s$^{-1}$ based on the
pseudo-bolometric light curves for SN 2007bg of \citet{2010A&A...512A..70Y}. Given the ratio
$U_{\rm{ph}}/U_{B}$ and the radio flux of $\nu F_{\nu}\approx2\times10^{-16}$ erg s$^{-1}$
cm$^{-2}$ the derived X-ray luminosity of $\sim2\times10^{-17}$ erg s$^{-1}$ cm$^{-2}$ is in
agreement with our upper limits. For later epochs we can not estimate the bolometric luminosity of
SN 2007bg.

\section{Summary and Conclusions}
\label{s:discussion}

In this article we presented the radio light curves and X-ray observations of Type Ic-BL SN 2007bg.
The radio emission is characterised by three different phases. Phase 1 being characterised by a SSA
turn-on. 
Based on the brightness temperature of SN 2007bg the expansion speed of the radio emitting region is
$v=0.19\pm0.02c$. An expansion speed similar to what has been found in the small sample of
radio-emitting Type Ib/c SNe.

The derived mass-loss rate from the progenitor during phase 1 is
$\dot{M}\approx1.9\times10^{-6}(v_{w}/1000\;\rm{km\,s}^{-1})$ \Msunpyr, a low mass-loss
rate among the observed in WR stars.
During phase 2 we observe a drop in the flux density while the spectral index seems to follow a
smooth evolution. This drop is presumably due to the radio ejecta entering a region of lower CSM
density, probably arising because of a change in the mass-loss rate of the progenitor or a variable
wind speed.
Then during phase 3 we observe a rise in the radio flux density along with a re-absorption of the
radio emission. The shallow observed slope of the re-absorption could be explained by small
clumps of material in the progenitor wind which would become more strongly self-absorbed than the
surrounding CSM when shocked and produce a shallower optically-thick spectrum. From our modelling
of the radio light curves during phase 3 we put an upper limit on the mass-loss rate of
$\dot{M}\la4.3\times10^{-4}(v_{w}/1000\;\rm{km\,s}^{-1})$ \Msunpyr.
This second wind component could arise from a different stellar evolution phase of the
progenitor prior to explosion, possibly a LBV phase. Or it could be the effect of stellar rotation
on the stellar wind properties.
If the radio flux density from SN 2007bg declines in time like $f_{\nu}\propto t^{-1}$. It should
still be detectable, with a flux density at $22.5$ GHz of $\sim500$ \mJy, and higher at lower
frequencies. If the progenitor star went through more mass-loss episodes, then the observed flux
density would deviate from this value.
Recent observations taken with the Karl G. Jansky Very Large Array (JVLA) should help to establish
this. Very few broad-lined type Ic SNe have radio detections, and none of them show signs of a
complex CSM like that of SN 2007bg.

These different wind components, along with the derived mass-loss rates will eventually help to
constrain the progenitors of Type Ib/c SNe, from their inferred pre-SNe properties. Currently two
main evolutionary paths has been proposed to account for the rate of these explosions: a single WR
origin, and a binary system. However, the single WR scenario itself still needs significant
clarification. For instance, what effects do evolution (mass-loss, rotation, etc.) and metallicity
have in shaping the CSM around massive type Ib/c progenitors and ultimately their explosions. And
more generally, how does this fold into the connection between type II's, type Ib/c, and GRBs.

Compounded by their relative rarity, a key impediment to improving our knowledge of broad-lined Ic
SNe has been the lack of high-quality, multi-frequency observations from early through late epochs.
Sensitive high-frequency radio observations of Ic SNe within 1-2 days of explosion are needed to
efficiently identify the best objects for follow-up and allow further refinement of the physical
properties surrounding these unique explosions. Both ALMA and the newly retooled JVLA will play
critical roles here, providing rigorous probes of the synchrotron-emitting shock (to constrain
evolution of the shock velocity, any potential beaming, etc.) and ultimately new insights
into the properties of their CSM (such as density, variability, clumpiness, and perhaps even dust
content for these massive systems). Sensitive, long-term X-ray and optical/NIR spectroscopic
follow-up are also needed in order to break potential degeneracies and provide consistency checks
against the shock-CSM interaction models. If such complete datasets can be assembled in the next
several years, they may afford us a leap in our physical understanding of SNe and GRBs.

\section*{Acknowledgments}

We thank the VLA for regularly monitoring this object, the VLA archive for making this data publicly
available, A. M. Soderberg for her telescope time, to M. Krauss for her help, 
K. Stanek and S. Schulze for their useful comments while drafting this article.
We acknowledge support from Programa de Financiamiento Basal (F. E. B.), the Iniciativa Cient\'ifica
Milenio through the Millennium Center for Supernova Science grant P10-064-F (F. E. B.), and
CONICYT-Chile under grants FONDECYT 1101024 (P. S., F. E. B.), ALMA-CONICYT 31100004 (F. E. B.), and
FONDAP-CATA 15010003 (F. E. B.), J. L. P. acknowledges support from NASA through Hubble
Fellowship Grant HF-51261.01-A awarded by STScI, which is operated by AURA, Inc. for NASA, under
contract NAS 5-2655.

\bibliographystyle{mn2e}

\label{lastpage}

\end{document}